\documentclass[fleqn,usenatbib]{mnras}

\usepackage{newtxtext,newtxmath}

\usepackage[T1]{fontenc}

\DeclareRobustCommand{\VAN}[3]{#2}
\let\VANthebibliography\thebibliography
\def\thebibliography{\DeclareRobustCommand{\VAN}[3]{##3}\VANthebibliography}

\usepackage{tablefootnote} 
\usepackage{threeparttable} 

\usepackage{graphicx}	
\usepackage{amsmath}	

\title[Outward Migration in Nascent Stellar Groups]{Outward Migration in Nascent Stellar Groups}

\author[Steven W. Stahler]{
Steven W. Stahler,$^{1}$\thanks{E-mail: stahler@astro.berkeley.edu}
\\
$^{1}$Berkeley Astronomy Dept, U. of California, Berkeley, CA 94720, USA\\
}

\date{Accepted XXX. Received YYY; in original form ZZZ}

\pubyear{2023}

\begin{document}
\label{firstpage}
\pagerange{\pageref{firstpage}--\pageref{lastpage}}
\maketitle

\begin{abstract}
As a stellar group forms within its parent molecular cloud, new members first appear in the deep interior. These overcrowded stars continually diffuse outward to the cloud boundary, and even beyond. Observations have so far documented only the interior drift. Those stars that actually leave the cloud form an expanding envelope that I call the ``stellar mantle." Simple fluid models for the cloud and mantle illustrate their basic structure. The mantle's expansion speed is subsonic with respect to the cloud's dynamical temperature. I describe, in qualitative terms, how the expanding mantle and Galactic tidal radius might together shape the evolution of specific types of stellar groups. The massive stars in OB~associations form in clouds that contract before extruding a substantial mantle. In contrast, the more slowly evolving clouds forming open clusters and T~associations have extended mantles that encounter a shrinking tidal radius. These clouds are dispersed by internal stellar outflows. If the remaining group of stars is gravitationally bound, it appears as a long-lived open cluster, truncated by the tidal radius. If the group is unbound, it is a late-stage T~association that will soon be torn apart by the tidal force. The ``distributed" populations of pre-main~sequence stars observed in the outskirts of several star-forming regions are too distant to be stellar mantles. Rather, they could be the remnants of especially low-mass T~associations.
\end{abstract}




\section{Introduction}



This paper concerns the general problem of stellar groups in formation, with emphasis on the most common types in the Galaxy. While stars in the plane exhibit a continuum of surface densities \citep{B10}, it is possible, and useful, to identify the three groups that appear in greatest abundance.\footnote{In this paper, I apply the term ``group" to a generic aggregate of stars. I restrict ``cluster" to gravitationally bound groups, and ``association" to unbound ones.} In order of increasing $N$, the population per system, these are: T~associations (\hbox{${\rm log}\,N\approx 1-2$}), open clusters (\hbox{${\rm log}\,N\approx 2$}), and OB~associations (\hbox{${\rm log}\,N\approx 3$}). \citep[For compilations of relatively nearby examples in each category, see][Chap.~4.]{S04} Consider, in broad terms, the forces at play in any group. An isolated, purely stellar system, such as a mature open cluster, is held together by self-gravity. Opposing this cohesive force is the inertia of the stars themselves, i.e., their velocity dispersion. As long as this velocity dispersion is isotropic, it is convenient, and justifiable theoretically, to picture it as a dynamical temperature. Multiplying by the mass density yields a stellar pressure, whose gradient 
at every location opposes gravity to stabilize the system \citep[][Chap.~4]{B11}.

At earlier times, the group's parent molecular cloud plays an equally
important, or even dominant, role. The cloud, too, is self-gravitating. The two components of the system, stars and gas, are subject to the gravitational field generated by both. The gas also experiences an outward, resistive force. This is the gradient, not of the kinetic gas pressure of atoms and molecules, but of the effective pressure associated with the bulk, turbulent motion of the gas, which itself is threaded by the interstellar magnetic field. Observations tell us that this nonthermal motion permeates all clouds that actually spawn stellar groups \citep[e.g.,][Chap.~4]{S04}.

In a sense, the stars and gas behave like two chemical components in a planetary atmosphere, since both are subject to a common gravitational potential. A key distinction is that their ``partial pressures" are not
additive. The gas is supported against gravity only by its own turbulent pressure. Likewise, stars are supported only by the pressure associated with their velocity dispersion. In addition, the cloud is immersed in a large-scale, gaseous medium with its own finite pressure. The actual force bearing down on the cloud surface is usually minor compared to self-gravity, but the fact of the confinement itself is critical. Individual stars can cross the boundary between the cloud and its external medium; parcels of cloud gas cannot. The central thesis of this paper is that such migration of the stars does in fact occur. Stars drifting out of the cloud form an extended envelope that I call the ``stellar mantle." After motivating the existence of the mantle, I discuss its basic properties in a schematic fashion. I then sketch its possible role in the later evolution of a stellar group.

I begin, in \S II below, by reconsidering and justifying the very notion of a parent cloud for stellar groups, in light of recent observations. I then present the case for an outward drift of the newborn stars. Section~3 gives several illustrative examples of a cloud-mantle system. After deriving structural equations in spherical geometry, I solve them numerically. In \S IV, I conjecture how these findings fit into a larger, evolutionary picture that includes the Galactic tidal force. Finally, \S V compares this investigation with others in the literature, including previous observations of extended stellar halos in star-forming regions.

\section{Cloud and Mantle}
\subsection{Identity of Clouds}

The idea that stellar groups form in discrete clumps or clouds (interchangeable terms in this paper) was a natural one before more detailed spatial morphology of the star-forming gas became available. Within recent years, millimeter
observations, especially from the Herschel satellite, have supplied much more information. We now see
that denser molecular gas is permeated by a network of elongated filaments. Embedded within these
structures are the cores long known to produce individual stars. \citep[For a review, see][]{A14}. 
The pattern of filaments themselves varies widely. In the Taurus region, they are relatively well-ordered \citep{G08}. Upon closer examination, each of these main structures contains features that may themselves be filamentary \citep{T15}. In contrast, the filaments in the Aquila Rift Cloud appear crowded and tangled \citep{A10}.

Many stellar groups form at the junctions of intersecting filaments, known as hubs \citep{M09}. Indeed,
almost all the infrared-detected, embedded groups in the Rosette Molecular Cloud coincide, in projection,
with hubs \citep{S12}. It might appear, then, that these hubs have supplanted the traditional clumps. Note,
however, that even a hub can be complex internally. In the Mon~R2 region, \citet{K22} found that the
most massive hub is not monolithic, but contains many relatively short, higher-density filaments.

As newer observations accrue, it remains true that the highest concentration of young stars coincides
with the densest molecular gas, whatever its detailed structure. Away from this central region, one still
may detect filaments. However, these tend to have lower masses per unit length, and fewer of them contain dense cores, even starless ones \citep[e.g.][Fig.~6]{M16}. In a study such as the present one, aimed at clarifying the main processes underlying stellar group formation, it is sensible to perform an ``average over filaments," and concentrate on the larger-scale structure. It is also true that each star-forming region has interacted from birth with its environment, and may still be accreting gas even as it forms stars \citep{G11}. In the same spirit of paring down to essentials, we may ignore such flows, under the assumption that they join the cloud far from the most active region of star formation. Note that a different picture may apply in the formation of young massive clusters, which we do not consider here. In these, even more populous, systems \hbox{(${\rm log}\,N\approx 4$)}, gas may coalesce and simultaneously form stars, in ``conveyer belt" fashion \citep{L14}. To the extent that this picture is further substantiated, continual gas accretion probably plays a role even in lower-mass systems, as anticipated by \citet{K20}.

\subsection{Basic Properties}

Individual clouds, whether relatively isolated entities or clumps within giant complexes, are the sources of the three principal stellar groups in the Galaxy. However, a specific group may not originate in a single cloud. Consider OB associations, the most massive of the three types. Most of these are gravitationally unbound, and therefore expanding. However, it has been difficult to pinpoint a unique geometric center for their outward proper motions. \citet{M17} managed to trace back the proper motion vectors in Per~OB1 and Car~OB1, and found minimum (and thus presumably initial) sizes for both in the range $11 - 27~{\rm pc}$. These sizes are characteristic of giant complexes as a whole, rather than their internal clumps \citep{S85}. Thus, there are probably multiple expansion centers, as \citet{Q22} recently discovered in Cygnus. For OB associations, at least, numerous clumps spawn the presently observed stars.

\citet{B07} have reviewed the observational data and the inferred physical properties of optically dark clumps. Technically, they are identified as coherent regions in position-velocity space \citep{W94}. In any 
sufficiently large complex, the majority of clumps have relatively low mass and are not self-gravitating, but 
are confined by the external pressure. However, most of the total system mass resides in 
the minority population of especially high-mass clumps.
Table~1 of \citet{B07} lists a clump mass range of $50-500~M_{\sun}$ and a corresponding 
range in size (equivalent circular radius) of \hbox{$0.3-3\,\,{\rm pc}$}. Note that this tabulation does not include 
the even more massive infrared dark clouds. which spawn both populous clusters and high-mass stars \citep{G09}. 

All these clouds have interiors in a state of supersonic motion. This situation is in marked contrast with that in
filaments and their embedded cores, whose line widths indicate essentially thermal speeds \citep{H11}. The effective pressure from internal motion in larger clouds counteracts the compressive force of self-gravity, as well as the pressure of the external medium. Analysis of the clouds' spectral line emission verifies that the motion is similar to classical turbulence, in that energy cascades from larger to smaller eddies \cite[e.g.,][]{H05}. One source for this turbulence is molecular outflows from young stars \citep{K00}. On the other hand, some clouds of comparable size, but with little detectable star formation, have similar internal motion \citep{M85}. Thus, there must also exist ``primordial" turbulence \citep{L06}, whose origin is plausibly gravitational contraction of the cloud \citep{G20}.

Since the largest-scale internal speeds are comparable to the Alfv\'en velocity \citep{C12}, the turbulence is magnetohydrodynamic. This MHD turbulence pervades the interior of the cloud, probably at a level that has only minor spatial variation. Evidence for this uniformity is suggestive, and certainly not ironclad. \citet{Ki19} found a relatively small spatial gradient in the velocity dispersion of young stars in the rich Orion Nebula Cluster. In addition, a plausible and empirically supported assumption is that the velocity dispersion of stars, once free of their dense cores, matches that of the parent cloud gas, in regions where both are present \citep[e.g.,][]{R15}. This equality does not hold for stars still forming inside their dense cores. The core-to-core velocity dispersion is lower than that of surrounding gas \citep{F15}, presumably because the cores are still locked within filaments. 

A concise theoretical view of the situation is that a cloud which is forming a stellar group is an isothermal structure, whose equivalent temperature is the common velocity dispersion of stars and gas. Another key, empirical fact is that the rate of star formation in regions still containing gas is accelerating \citep{C18}. Such acceleration indicates that the parent clouds are contracting, thus driving up their density. The morphology of both the clouds and their newborn groups indicates that this contraction is unlikely to be true, free-fall collapse, but is instead quasi-static in nature \citep{T06}. The contraction rate is set by the dissipation of internal turbulence, which occurs in low-velocity shocks (see \S~3.1 below). As the cloud shrinks, its equivalent temperature evolves, in order to maintain near force balance. 

\subsection{Stellar Drift}

Returning to the more detailed findings, the densest region of the parent cloud is crowded with filaments of relatively high mass per unit length. These, in turn, contain numerous cores that themselves produce stars. The general correlation between stellar and gas density has long been known. \citet{P02} showed that the visible T~Tauri stars in Taurus are mostly located in or near the main filaments, as delineated in CO. More recently, \citet{J08}, in their study of Ophiuchus, found that mid-infrared stellar sources are closely associated with peaks in submillimeter continuum emission, where the latter stems from dust grains embedded in the molecular gas. \citet{S16} found the equivalent result in the Orion~A star-forming region.

Another trend that has come to light is that the correspondence of stellar and gas densities is tightest for the least evolved stars. \citet{J18} found that about half of Taurus stars belong to 20 cohesive groups. The location of these groups is well correlated with the gas column density, and the groups themselves are rich in Class~I sources. \citet{G22} plotted the locations of 269 young stars in Ophiuchus, and compared their distribution with that of peaks in 850~$\mu$m emission. For the youngest, Class~0 and I, sources, the peak of the stellar distribution coincides with that of the emitting dust. Class II and III sources have a noticeably flatter distribution, extending to several tenths of a parsec beyond the nearest dust peaks, well beyond their filaments of origin. A similar pattern is evident in observations of the integral-shaped filament of Orion~A \citep{S16}. In summary, stars do not remain in situ once formed, but drift outward from their birthsites.

This drift stems from an overdensity of the very youngest stars. Adopting once again the simplfied theoretical perspective, all the stellar birthsites lie within the central, highest-density portion of the isothermal cloud. Such a spatial concentration of star formation activity has, in fact, long been observed \citep{L92}. It may be viewed as reflecting a local version of well-studied formation laws, such as the Schmidt-Kennicutt relation \citep{K98}. The latter, first derived for galaxies, is known to break down at the sub-kpc scale \citep[e.g.,][]{O10}. Nevertheless, an analogous relation plausibly holds {\it inside} parent molecular clouds. If so, the stellar mass produced per time and per volume varies as $\rho_g^N$. Here, $\rho_g$ is the mass density of molecular gas, and \hbox{$N\approx 1-2$}. \citet{H07} used a contracting spherical cloud to model the star formation history of the Orion Nebula Cluster. By adopting a relation of the above form, with $N\,=\,2.4$, they reproduced the observed acceleration of star formation in the group. 

The essential lesson is that stars are born with a more peaked density than their background gas. But observed young stars that are no longer deeply embedded in their dense cores do not exhibit this difference. (Compare, e.g., the distribution of Class~I and II sources in \citet[][Fig.~10]{G08}.)
The implication, again, is that stars cannot remain in place after they form, but must migrate outward. Meanwhile, new stars, similarly crowded, take their place, and the drift persists throughout the cloud. Of course, the cloud itself evolves with time, but the observations indicate that outward stellar drift dominates this background motion, at least during the epoch of greatest star formation activity.

Observations provide the means to estimate, at least roughly, the actual drift speed. According to Figure~4 of \citet{G22}, Class~II sources in Ophiuchus have a mean separation from their respective 850~$\mu$m peaks that is 0.04~pc greater than the mean separation for Class~0 and I sources. Transforming this projected distance into a three-dimensional one, and taking the Class~II lifetime to be 2~Myr \citep{E09}, we find a drift speed of 0.03~km~s$^{-1}$. If we repeat this exercise using only those stars in both stages that have the greatest separation from the gas, the estimated speed increases to 0.08~km~s$^{-1}$. Both figures, as we shall confirm, are well under the cloud's own velocity dispersion. It may be that the drift speed across the cloud boundary is greater (see \S~3.5 below), but young stars still migrate from their birthsites at a subsonic rate.

What drives the stars outward? The stars peak in density near their birth region, so it is tempting to ascribe their motion as the response to excess stellar pressure. Within a diffusion picture, it is the fall in pressure that creates outward motion. To justify this description, we should check that there is actually time for the stars to interact significantly, i.e., to perturb each others' orbits. Such perturbations require a relaxation time. In a system of $N$ stars, this is a factor {$f\,=\,0.1~N/{\rm ln}\,N\,$} larger than the dynamical crossing time \citep[][Chap.~1]{B11}. For the populations of main interest \hbox{($N\,\sim\,10^2-10^3$)}, we have \hbox{$f\,\sim\,3-14$}. It is doubtful that most clouds are forming stars for more than a few crossing times \citep{T06}. On the other hand, the relevant crossing time is not that for the entire cloud, but only for the interior portion actively forming stars. This time is at most a few Myr. In addition, the most populous groups disperse before significant drift, as discussed below. The outward migration, when it occurs, therefore does arise effectively from stellar pressure. The excess is continually regenerated, as long as star formation persists. Stars spread out so that their distribution is appropriate for the velocity dispersion and the gravitational potential of the system, both of which evolve with time.

\subsection{Galactic Tide}

A final influence on the cloud and its stars is an external one, the tidal component of the Galactic gravitational field. This force plays a role, not in the birth of stellar groups, but in their demise. Any loosely bound cluster or dwarf spheroidal galaxy that comes near enough to the Galactic center is quickly ripped apart \citep{O95}. At our location, the tidal force acts more slowly, gradually eroding an initially massive cluster, and setting its total life span \citep{L05}. Lower-mass open clusters again disrupt rapidly, from the tidal shock created by a passing giant molecular cloud \citep{T87}.

Even when it does not destroy a group entirely, the Galactic tidal force establishes an effective outer boundary. This tidal radius, as first obtained by \citet{K62}, is
\begin{equation}
R_T \,=\, \left[ \frac{G\,M}{4\,A\,(A-B)}\right]^{1/3}  \,\,.
\end{equation}
Here, $M$ is the cluster mass, while $A$ and $B$ are the Oort constants. The radius $R_T$ is the distance of the (unstable) Lagrange points L1 and L2 from the center of the cluster. Stars that wander outside this critical radius are lost to the field. Thus, the stellar density in any 
sufficiently extended cluster falls sharply in the vicinity of $R_T$.

This truncation is well documented for globular clusters, whose actual radii are consistent with the tidal values, given a plausible mass-to-light ratio \citep[e.g.,][]{B04}. For open clusters, the tidal and physical radii are similar, but pinpointing the latter is problematic, owing to the systems' relative sparseness; researchers are currently employing GAIA to make headway \citep{C21}. Meanwhile, \citet{P07} analyzed a catalog of 236 open clusters, estimating their radii through a \citet{K66} model fit to their observed stellar density profiles. Assuming the radii were tidal in origin, the authors inferred system masses. For relatively nearby clusters, these were consistent with masses obtained through traditional star counting, after due extrapolation for incompleteness. 

Returning to the general topic of young stellar groups, there is no evidence that star-forming molecular clouds or clumps are spatially limited by the Galactic tidal field, much less destroyed by it. Thus, each cloud must sit well within its associated tidal radius. For a representative clump mass of $500\,\,M_{\sun}$, equation~(1) gives \hbox{$R_T\,=\,11\,{\rm pc}$}, about 4 times the typical, corresponding gas cloud radius of $R_g\,=\,3~{\rm pc}$ \citep{B07}. This separation of the tidal radius from that of the cloud must diminish with time, at least for those clouds destined to become open clusters.

What actually determines the size of a star-forming molecular cloud for most of its existence is internal turbulence, self-gravity, and the pressure of an external, gaseous medium, hotter and more rarefied than the cloud itself. The physical nature of this confining gas is still uncertain. \citet{W95} adduced evidence that the clumps within the Rosette giant molecular cloud, which are the sites of stellar group formation, are surrounded by HI gas. So too are the smaller clumps in the Perseus molecular cloud \citep{L12}. In any event, it is this relatively sparse gas into which the stellar mantle expands as its parent cloud continues to form stars. Expansion of the mantle continues until the denser, molecular gas eventually disperses to reveal a bare stellar group. 







\section{Quantitative Formulation}
\subsection{Dynamical Equations}

By ascribing to the ensemble of stars a local velocity dispersion, one is essentially treating them as a fluid. Many theorists in the past have solved stellar dynamical problems in this manner, notably in the study of globular clusters \citep[e.g.,][]{G87}. The fluid picture is technically valid as long as the mean free path for stellar interaction is sufficiently short, and the velocity dispersion is spatially isotropic \citep[][Chap.~4]{B11}. Note that existing N-body simulations can easily track individual stars in the populations of interest here. However, the fluid description is more appropriate in the present case, since the background potential is first dominated by the gas cloud and is unknown in advance. 

It is possible, then, to derive a set of fluid equations governing a contracting cloud that is forming stars. In this paper, I will not attempt to describe theoretically the full evolution of the star-forming cloud and its interior stars.  
Rather, I simply provide snapshots of the cloud and its mantle at a typical time. This much more restricted description requires only a selected, and simplified, subset of the equations. In any event, I begin by presenting the full set that any future, more complete, account will utilize.

Since these equations govern both gas and stars, I obtain them by taking moments of the respective Boltzmann equations. This approach to fluid dynamics from kinetic theory is a standard one, well described in \citet{S92}. However, the system here is nonstandard in several respects. First, I am not concerned with individual molecules of the gas, but rather macroscopic fluid elements, whose fluctuating velocities constitute the cloud's turbulence. In addition, this component of the system gradually transforms into the other, the stars. To accommodate this process, I start with modified Boltzmann equations. 

Finally, when considering energy, I make the convenient but clearly unrealistic assumption that gas fluid elements collide elastically. The actual dissipation takes place in the smallest turbulent eddies, and occurs in relatively low-velocity shocks, such as those analyzed by 
\citet{P12}. For present purposes, I do not require, and indeed cannot accommodate, this level of detail. Instead, I simply add to the gas energy equation a term for radiative loss. The interested reader may consult Appendix~A for the full derivation.

We begin with equations governing the cloud. The first two express mass continuity for the gas and stars. If $\rho_g$ and $\rho_s$ are their respective mass densities and $\boldmath{u}$ their common bulk velocity, then
\begin{align}
{{{\partial}\rho_g}\over{dt}} \,&+\, {\boldmath\nabla}\!\cdot\!\left(\rho_g{\boldmath{u}}\right) \,+\,s_\ast \,=\,0 \\
{{{\partial}\rho_s}\over{dt}} \,&+\, {\boldmath\nabla}\!\cdot\!\left(\rho_s{\boldmath{u}}\right) \,-\,s_\ast \,=\,0 \,\,.
\end{align}
Here, $s_\ast$ is the star formation rate, expressed as stellar mass produced per unit volume per time. If \hbox{$\rho\equiv\rho_g+\rho_s$} is the total mass density, then (2) and (3) sum to the more familiar relation
\begin{equation}
    {{\partial\rho}\over{dt}} \,+\,{\boldmath\nabla}\!\cdot\!\left(\rho{\boldmath{u}}\right) \,=\,0 \,\,.
\end{equation}

Momentum conservation for the two components is
\begin{align}
    {\partial{\boldmath u}\over{\partial t}} \,&+\,\left({\boldmath u}\!\cdot\!{\boldmath\nabla}
    \right){\boldmath u} \,+\,{1\over\rho_g} {\boldmath\nabla} P_g  \,+\, 
    {\boldmath\nabla}\psi \,=\, 0      \\
    {\partial{\boldmath u}\over{\partial t}} \,&+\,\left({\boldmath u}\!\cdot\!{\boldmath\nabla} 
    \right){\boldmath u} \,+\,{1\over\rho_s} {\boldmath\nabla} P_s  \,+\, 
    {\boldmath\nabla}\psi \,=\, 0 \,\,.
\end{align}
Here, $\psi$ is the gravitational potential. I have used $P_g\equiv\rho_g c^2$ and $P_s\equiv\rho_s c^2$ to denote the partial pressures, with $c$ being the common sound speed. The latter is assumed to be spatially constant, but varies with time.

Noticeably absent from the last two equations is $s_\ast$, the formation rate of stars. Doesn't their creation remove momentum from the gas flow? Indeed, it does. However, as long as the newborn stars share the local velocity of the gas, the latter experiences no reaction force. There is also no extra force on the stars, to which mass is added. The proof for both cases is contained in Appendix~A. Note that the present circumstance differs from that in a ``cluster wind," in which stars that are the collective source of the wind are injecting gas with zero net velocity, and are thus exerting an effective drag \citep{C00}. 

The final dynamical relation is the heat equation:
\begin{equation}
    \rho\,c^2\, {{D\phantom{t}}\over{Dt}}\,{\rm ln}\left({c^3\over\rho}\right)
    \,+\,{\boldmath\nabla}\!\cdot\!{\boldmath F} \,=\,0\,\,,
\end{equation}
which includes both the convective time derivative 
\begin{equation}
  {D\phantom{t}\over{Dt}}\,
   \equiv\,{\partial\phantom{t}\over{\partial t}} 
    \,+\,{\boldmath u} \!\cdot\! {\boldmath\nabla}\,\,,
\end{equation}
and the specific entropy, ${\rm ln}\,(\,c^3/\rho)$. The energy flux 
$F$ has two components:
\begin{equation}
    {\boldmath F} \,=\, {\boldmath F}_{\rm mech} \,+\, {\boldmath F}_{\rm rad} \,\, .
\end{equation}
The first righthand term represents the mechanical transport of the turbulence in gas and stars. The second is the radiative flux, stemming from the gas alone. Only  mechanical transport exists within the stellar mantle. The turbulent flux is injected at the cloud boundary, and must vanish at the far edge of the mantle. Thus, there is a shift of the injected energy within the mantle, but no net gain or loss. In the present study, I do not track this internal shift, but simply assume that specific entropy is conserved. In other words, the mantle expands adiabatically. 

\subsection{Quasi-Static Approximation}

We have noted that the morphology of young stellar groups suggests they were born in a near-equilibrium condition. The implication is that their parent clouds were also in a state close to force balance. Self-gravity naturally induces contraction in these clouds, but the motion is apparently subsonic with respect to the dominant turbulent velocity. A similar conclusion follows from the earlier work of \citet{P00}, who found a relatively large stellar age spread in a number of star-forming regions. While there still is not full agreement in this matter \citep[see][for a contrary view]{K15}, I will henceforth adopt the slow-contraction hypothesis. A practical consequence is that, in our dynamical equations, we can ignore all explicit time derivatives. 

Recall that our present goal is to provide a snapshot of a typical cloud and its mantle. Within the cloud, we may also ignore the acceleration of both fluids, so that momentum balance is simply
\begin{align}
    {c^2\over\rho_g} {\boldmath\nabla} \rho_g  \,+\, 
    {\boldmath\nabla}\psi \,=\, 0      \\
    {c^2\over\rho_s} {\boldmath\nabla} \rho_s  \,+\, 
    {\boldmath\nabla}\psi \,=\, 0 \,\,.
\end{align}
Combining these two equations, we see that
\begin{equation}
    {\boldmath\nabla}\,{\rm ln}\,\rho_g  \,=\, {\boldmath\nabla}\,{\rm ln}\,\rho_s   \,\,.
\end{equation}
It follows that the gas and stellar density profiles inside the cloud are identical, apart from a proportionality constant. Equivalently, we define the mass fraction in stars as \hbox{$\chi\,\equiv\,\rho_s/\rho$}. As was true for the sound speed $c$, this quantity is spatially constant, but is generally a function of time. We stress that the density proportionality applies only within the parent cloud proper, i.e., that portion of the molecular gas that can potentially form stars; see also \S~3.4 below.

The foregoing result follows if both gas and stars are subject only to their collectively produced gravity. If, on the other hand, the stars exert an additional force on the gas, then the proportionality breaks down. Such a force could arise even from low-mass stars, which generate jets and molecular outflows that impart momentum to the gas. In addition, high-mass stars have powerful winds, and will eventually ionize their parent clouds. All these processes play a role in driving off the gas and revealing a bare stellar group. We restrict our snapshot, then, to times prior to the final gas dispersal.

In the derivation of the dynamical equations, it is assumed that star formation occurs quickly, in comparison to the cloud's evolutionary time scale. We also noted that the star formation rate $s_\ast$ is more centrally peaked than the gas. Yet the combined momentum equations imply that the stellar density profile parallels that of the gas. This flattening of the stars' spatial distribution is a manifestation of their outward drift. Another implicit assumption, that the drift speed is subsonic, is corroborated by the limited observations available. 

\subsection{Equations in Spherical Geometry}

Consider the idealized case of a spherical parent cloud. If $r$ is the distance from the cloud center, then the gravitational potential is
\hbox{$\psi\,=\,-G M_r/r$}. Here, $M_r$ is the mass interior to any radius:
\begin{equation}
    M_r \,=\,\int_0^r \! 4\,\pi\,r^2\rho\,\,dr \,\,. 
\end{equation}
Both in the cloud interior and the mantle, we solve the equations for hydrostatic balance: 
\begin{align}
    {{dP}\over{d r}} \,&=\,-\rho\,{{G M_r}\over r^2} \\
    {{d M_r}\over{dr}} \,&=\, 4\,\pi\,r^2\rho \,\,. 
\end{align}
Within the interior, \hbox{$P\,=\,\rho\,c^2$}, so equation~(14) becomes
\begin{equation}
    {{d \rho}\over{d r}} \,=\,-\rho\,{{G M_r}\over{c^2\,r^2}}\,\,.\qquad{\it cloud}\,\,{\it interior} 
\end{equation}
We find the gas density through 
\hbox{$\,\,\rho_g = (1-\chi)\,\rho$},
and the stellar density through \hbox{$\,\,\rho_s = \chi\,\rho$},
where $\chi$ is a constant. Since there is no central point mass, \hbox{$M_r(0)\,=\,0$}. At the boundary of the cloud, we set the gas pressure, $\rho_g c^2$, equal to $P_e$, the pressure in the external medium.

Inside the stellar mantle, we continue to solve equations~(14) and (15), with the total density now being the stellar one: \hbox{$\rho\,=\,\rho_s$}. The pressure and density are related by the adiabatic relation:
\begin{equation}
    P \,=\,P_{\rm se} \left( \rho\over \rho_{\rm se} \right)^{5/3} \,, \qquad{\it stellar}\,\,{\it mantle}
\end{equation}
where $P_{\rm se}$ and $\rho_{\rm se}$ are the stellar pressure and density, respectively, at the cloud edge. Thus, these quantities are assumed to be continuous across that interface. Both the cloud and its mantle lie within the Galactic tidal radius, $R_T$. The latter depends on $M_{\rm tot}$, the total mass of gas and stars, following equation~(1).

The mantle is actually a flow of stars, extruded by the cloud. Stars experience an outward force associated with their pressure gradient. Thus, the mantle constitutes a thermal wind. Even if such a wind starts subsonically, it typically accelerates to the local sound speed and beyond. However, an adiabatic thermal wind, with \hbox{$\gamma\,=\,5/3$},
remains gravitationally bound to the central source, and only decelerates \citep[][Chap.~4]{L99}. Our simpler model of a static envelope is thus adequate.

I solve the system of equations by transforming to nondimensional variables. Here, I take $G$, $P_e$, and $c$ to be fiducial quantities. Using a tilde to denote the new, nondimensional quantities. I set
\begin{align}
    \tilde r \,&\equiv\,{{(4\pi G P_e)^{1/2}\,r\over c^2}} \\
    \tilde\rho \,&\equiv\,{\rho\,c^2\over P_e} \\ 
    \tilde M_r\,&\equiv\, {{\left(4\pi P_e G^3\right)^{1/2}\!M_r}\over c^4}
\end{align}
The equations then become 
\begin{align}
    {{d {\tilde\rho}}\over{d{\tilde r}}}\,&=\,-{\tilde\rho}\,{{ {\tilde M}_r}\over{{\tilde r}^2}} 
    \ \ \ \qquad\qquad\qquad\qquad{\it cloud}\,\,{\it interior}\\
    {{d {\tilde\rho}}\over{d{\tilde r}}}\,&=\,-{3\over 5}\left({\chi\over{1-\chi}}\right)^{2/3}\!
    {\tilde\rho}^{1/3}\,{{ {\tilde M}_r}\over{{\tilde r}^2}} 
    \qquad{\it stellar}\,\,{\it mantle}\\
    {{d {\tilde M}_r}\over{d{\tilde r}}} \,&=\, {\tilde r}^2\,{\tilde\rho}  \,\,\,.
\end{align}

Let ${\tilde R}_g$ be the nondimensional cloud radius. This point is reached when the nondimensional gas density falls to unity, or when \hbox{$\tilde\rho (\tilde R_g) = 1/(1 - \chi)$}. Proceeding outward, the initial density in the mantle is \hbox{$\chi/(1-\chi)$}. In the present examples, integration continues until the mantle has a certain, specified mass. The tidal radius is expressed as 
\begin{equation}
    {\tilde R}_{\rm T} \,=\, \alpha_T\, {\tilde M}_{\rm tot}^{1/3}
    \,\,\,.
\end{equation}
Here, the nondimensional quantity $\alpha_T$ is
\begin{equation}
\alpha_T \,\equiv\, \left({{\pi\,G\,P_e}\over{c^2\,A\,(A-B)}}\right)^{1/3} \,\,.   
\end{equation}

Evaluation of $\alpha_T$ requires adopting representative values for $c$ and $P_e$. For the first quantity, we may use the line widths of optically thin tracers. To cite one representative study, \citet{L89b} mapped the Ophiuchus region extensively in $^{13}$CO. His Table~3 gives, for the main, L1688 cloud, a one-dimensional (radial) velocity dispersion of 0.88~km~s$^{-1}$. Multiplying by $\sqrt{3}$ for the three-dimensional value yields \hbox{$c\,=\,1.5\,\,{\rm km}\,\,{\rm s}^{-1}$}. 

Assigning a value for $P_e$ is more problematic, as the nature of the medium bounding star-forming clouds remains uncertain. Assuming it to be HI gas within the Rosette Molecular Cloud, \citet{W95} estimated, from observations of the relevant density and velocity dispersion, \hbox{$P_e\,=\,5\times 10^4\,\,{\rm K}\,\,{\rm cm}^{-3}$}. \citet{E89} arrived at essentially the same figure, starting from the observed surface density. For comparison, \citet{C05} gives \hbox{$\,2\times 10^4\,\,{\rm K}\,\,{\rm cm}^{-3}$} as the partial pressure of HI in the general interstellar medium. It is fortunate, in any case, that $\alpha_T$ is insensitive to the precise values of $P_e$ and $c$. Inserting the above estimates yields \hbox{$\alpha_T\,=\,5.3$}. 

\subsection{Illustrative Examples}

Integration of the equations requires first assigning a value to $\chi$. We note parenthetically that this quantity would appear, as one result among many, in a full evolutionary calculation. Pending such an endeavor, I adopt $\chi$ from observations, where it is traditionally known as the ``star formation efficiency." Surveying the literature, one finds a broad range in quoted values. In part, this spread represents true variation in different star-forming clouds. Another source of discrepancy is ambiguity in the very definition of a parent cloud. As the minimum density considered for this entity is raised, so that the associated cloud mass shrinks, the inferred efficiency is greater. There is also the potential problem of including too much mass. It is now appreciated that most molecular gas generally is of such low density that it may never participate in the star formation process; \citet{E21} have recently stressed this point.

With these difficulties in mind, we consider two examples, both in well-studied locales, which yield a plausible range of $\chi$. Within the Taurus association, the L1495 and B209 regions are adjacent sites of star formation. \citet{L23} has identified 70 member stars, according to their spatial clustering and similar proper motions (see his Fig.~1 and Table~2). The combined gas mass of the regions, from Herschel-derived column densities, is 1500~$M_\odot$ \citep{M16,A23}. Assigning each star a nominal mass of 0.5~$M_\odot$ yields $\chi\,=\,0.02$. 

In Ophiuchus, the most prolific star-forming cloud is L1688, whose estimated mass, from $^{13}$CO observations, is 1800~$M_\odot$ \citep{L89a}. 
The total enclosed stellar mass is $92\,M_\odot$ \citep{B01}, so that $\chi\,=\,0.05$. It is interesting to note that Ophiuchus is actually younger in terms of star formation activity than Taurus \citep{P00}. The higher $\chi$-value may indicate that its parent clouds were denser even before they began to produce stars. Indeed, their morphology suggests shock compression by the nearby Sco~OB2 association \citep{L86}. In any event, I adopt, as a representative range, \hbox{$\chi\,=\,0.02-0.05$}. The upper bound agrees with \citet{T02}, who derived cloud masses exclusively from $^{13}$CO data.

Integration of equations~(21)-(23) finally requires assigning a starting density ${\tilde\rho}(0)$. It is more meaningful physically to choose the initial, nondimensional gas density, ${\tilde\rho}_g(0)$, which corresponds to the cloud's center-to-edge density contrast. A lower limit is the critical Bonnor-Ebert contrast of 14. This characterizes, in an isothermal sphere, that configuration for which self-gravity alone first counteracts internal pressure \cite[][Chap.~9]{S04}. Inserting \hbox{${\tilde\rho}_g(0) = 14$} and \hbox{$\chi = 0.02$}, we integrate the equations out to ${\tilde R}_m$, which we select as the radius where the mantle mass equals that of the stars inside the cloud. We find that ${\tilde R}_m/{\tilde R}_g\,=\,1.7$. As anticipated, the tidal radius is much further out, ${\tilde R}_T/{\tilde R}_g\,=\,5.0$. Adopting the maximum $\chi-$value of 0.05, these figures change only slightly, to
${\tilde R}_m/{\tilde R}_g\,=\,1.7$ and ${\tilde R}_T/{\tilde R}_g\,=\,5.1$.

It is more likely that clouds do not produce copious stars until they are more strongly self-gravitating, i.e., until their density contrasts have surpassed the critical Bonnor-Ebert value. An isothermal sphere can be in a dynamically stable, equilibrium state for much higher contrasts, up to \hbox {${\tilde\rho}_g (0) = 380\,$} \citep{C03}. (The Bonnor-Ebert state is a stability transition only when considering relatively slow, isothermal perturbations.) As a further example, we use \hbox {${\tilde\rho}_g (0) = 50$}. The previous results do not change substantially. For 
\hbox{$\chi = 0.02$}, we find ${\tilde R}_m/{\tilde R}_g\,=\,1.9$ and ${\tilde R}_T/{\tilde R}_g\,=\,5.5$. For \hbox{$\chi = 0.05$}, only the tidal radius changes, to ${\tilde R}_T/{\tilde R}_g\,=\,5.6$. These results are summarized in Table~1. In addition, Figure~1
displays both the gas and density profiles for the last case. Within the cloud, both profiles tend toward power laws, with an exponent of $-2$, as is generally true for isothermal spheres. The mantle density also descends in nearly power-law fashion, but the slope is shallower, as expected for adiabatic expansion. 
\begin{table}
	\centering
	\caption{Numerical results for clouds of varying star formation efficiency and density contrast. The quantities ${\tilde V}_m$ (expansion speed of mantle) and ${\tilde V}_d$ (drift speed across cloud boundary) are upper bounds.}
	\label{tab:example_table}
	\begin{tabular}{lccccr} 
		\hline
		$\chi$ & ${\tilde\rho}_g(0)$ & ${\tilde R}_m/{\tilde R}_g$ & ${\tilde R}_T/{\tilde R}_g$ & ${\tilde V}_m$ & ${\tilde V}_d$ \\
		\hline
		0.02 & 14 & 1.69 & 5.03 & 0.34 & 0.92\\
		0.02 & 50 & 1.92 & 5.52 & 0.51 & 0.85\\
		0.05 & 14 & 1.71 & 5.13 & 0.18 & 0.34\\
        0.05 & 50 & 1.92 & 5.63 & 0.22 & 0.31\\
		\hline 
	\end{tabular}
\end{table}
\begin{figure}
    \hspace{-0.00truecm}\includegraphics[width=\columnwidth]
    {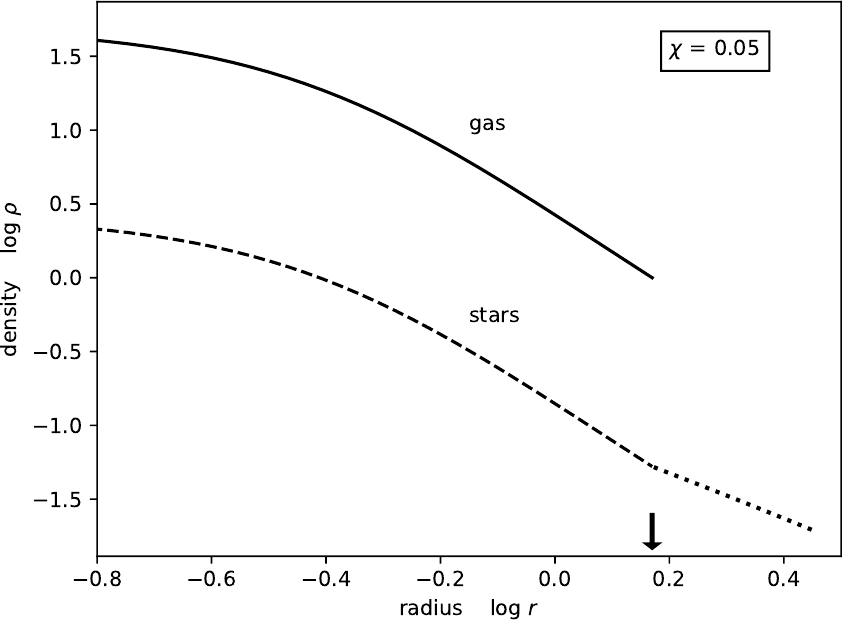}
    \caption{Gas ({\it upper curve}) and star ({\it lower curve}) density profiles, for $\chi = 0.05$ and ${\tilde\rho}_g(0) = 50$. Beyond the cloud boundary, indicated by the vertical arrow, the stars extend into the mantle. The mantle is assumed in this example to have the same mass as the stars inside the cloud.} 
\end{figure}
	
\subsection{Maximal Expansion Velocities}

How fast does the mantle spread? We may estimate the expansion speed by using our snapshots of representative mantle conditions. Reverting to dimensional notation, the outer edge expands at $V_m$, where 
\begin{equation}
V_m\,=\,{{dR_m}\over{dt}}\,=\,{{dR_m}\over{dM_m}}\,\,\,
{{dM_m}\over{dt}} \,\,.
\end{equation}
We obtain an upper limit to $V_m$ by assuming that {\it all} of the stars created inside the cloud feed the mantle. If $S_{\rm tot}$ is the total stellar mass produced per time, then we take
\begin{equation}
{{dM_m}\over{dt}}\,=\,S_{\rm tot} \,\,.
\end{equation}

Recent studies have found an empirical relationship between $S_{\rm tot}$, the mass of the cloud, $M_g$, and the cloud's free-fall time, $t_{\rm ff}$:
\begin{equation}
    S_{\rm tot} \,=\, \epsilon_{\rm ff}\,
    {M_g\over t_{\rm ff}} \,\,.
\end{equation}
The quantity $\epsilon_{\rm ff}$ is termed the ``star formation efficiency per free-fall time." 
\citet{K19} have provided a thorough review of the observations underlying this result, along with the associated uncertainties. According to these authors, most evidence favors $\epsilon\,\approx\,0.01$, the value I adopt.

After utilizing equation~(28), along with the conventional definition of $t_{\rm ff}$, the limiting, nondimensional  mantle speed is
\begin{equation}
    {\tilde V}_m \,=\,\epsilon_{\rm ff}\,\sqrt{8\over\pi^2}\,\,
    {{d{\tilde R}_m}\over{d{\tilde M}_m}}\,\, 
    \left({{\tilde M}_g\over {\tilde R}_r}\right)^{3/2} \,\,.
\end{equation}
Recall that \hbox{${\tilde V}_m\,=\,V_m/c$}. where $c$ is the velocity dispersion in the cloud. I evaluate the derivative in the above expression by numerically differencing the snapshot models. For the previous choices of $\chi$ and ${\tilde\rho}_g(0)$, ${\tilde V}_m$ varies from 0.2 to 0.5. Table~1 displays the full results. The mantle expands subsonically, as we might expect, given the relatively low drift speed of the stars that inflate it.

Returning to this internal drift speed, we may obtain an upper bound on ${\tilde V}_d$ at the cloud boundary, in the same spirit as our assessment of ${\tilde V}_m$. Assuming once more that all the stars being produced exit the cloud, the maximal speed is
\begin{equation}
    V_d \,=\, {S_{\rm tot}\over{4\,\pi\,R^2\,\rho_{\rm sb}}}
    \,\,,
\end{equation}
where $\rho_{\rm sb}$ is the stellar density at the cloud boundary. Again using equation~(28) for $S_{\rm tot}$, we find the nondimensional result
\begin{equation}
    {\tilde V}_d\,=\,{\epsilon_{\rm ff}\over{3\,n}}\,\, 
    {{\tilde\rho_{\rm av}}\over{\tilde\rho_{\rm sb}}} \,\,.
\end{equation}
Here, $\tilde\rho_{\rm av}\equiv 3\,{\tilde M}_g/{\tilde R}_g^3$ is the average gs density of the cloud. The number $n$ is the ratio of the free-fall to crossing time:
\begin{equation}
    n\,\equiv\,{c\,t_{\rm ff}\over R}\,=\,
    \sqrt{{\pi^2 {\tilde R}_g}\over{8\,{\tilde M}_g}}\,\,.
\end{equation}
This quantity is nearly identical for all our parameter choices, about 0.7. In any event, we may readily evaluate equation~(31) for each model considered. The result, shown in the rightmost column of Table~1, is that ${\tilde V}_d$
varies from 0.3 to 0.9. Dimensionally, these figures correspond to 0.5 and 1.4~km~s$^{-1}$, for our representative $c$-value of 1.5~km~s$^{-1}$. Comparison of this estimation with our earlier one in \S~2.3 suggests that the drift speed increases toward the outer region of the cloud. A similar trend has been noted by \citet{C15}, in their observational study of the young open cluster IC~348. In any event, the drift speed still appears to be at least marginally subsonic.

\section{Evolutionary Context}

\subsection{Role of Cloud Mass}

Thus far, we have not emphasized the actual mass of the parent cloud. beyond citing typical observed values. But higher-mass and lower-mass clouds evolve differently. Self-gravity is naturally stronger in more massive clouds. Like massive pre-main-sequence stars, which also contract quasi-statically through the release of internal energy, these clouds shrink relatively quickly. Conversely, lower-mass clouds take longer to reach sufficient density for star formation. In their study of the Rosette Molecular Cloud, \citet{W95} showed that only clumps which are manifestly self-gravitating contain embedded young stars. This finding is consistent with our previous remarks on the critical Bonnor-Ebert density contrast and its physical significance.

A number of researchers have studied the mass distribution of molecular clumps; \citet{B07} summarize their findings. The number of clumps per unit mass scales as $M^{-\alpha}$, where \hbox{$1.4 < \alpha < 1.8$}. The fact that \hbox{$\alpha < 2$} has the important consequence mentioned in \S 2.2. In any sufficiently large collection of clumps, most are of relatively low mass. On the other hand, the total mass of the aggregate is dominated by the relatively few with the greatest individual mass. 

There is a striking parallel between this finding and the statistics of stellar groups. It is widely appreciated that OB associations, the most populous and massive of the three types, contribute the majority of stars being formed at any time \citep[e.g.,][]{Le12}. Equally  instructive is the frequency with which each group is created within the Galaxy. Here, the classic study was by \citet{M78}. The specific numbers in this paper have long been superseded, but not enough to alter their overall ordering by production rate. According to these authors, T~associations, the {\it least} populous group, are created at by far the highest rate. The suggestion is that OB~associations form, at least in part, from the relatively rare clumps of greatest mass. T~associations are created by the much more common, low-mass clumps. Finally, open clusters originate from intermediate-mass clouds.

High-mass clumps alone are insufficient to account for OB~associations. Every such group originates in a giant molecular cloud, or complex, and each complex is pervaded by numerous clumps \citep{B93}. Most are not self-gravitating. An appreciable fraction, however, do contract and form stars, which comprise the bulk of the low-mass members in each association. Their parent clouds are not at the high end of the clump mass spectrum. It is only a few in this last category, in the tail of the distribution, that form massive O and B stars. 

Consider finally a clump that does form a massive star. Once created, the star's powerful wind and ionizing radiation destroy not only the clump, but the entire molecular complex. Prior to this event, the clump was contracting and forming many low-mass stars. These were centrally peaked in the clump, setting the stage for outward migration. However, the contraction itself was relatively rapid, so it is doubtful that the subsonic drift had time to create an extensive mantle. The velocity dispersion of the stars, both in this clump and others in the complex, was relatively high, reflecting the depth of the gravitational potential well. Following dispersal of the gas, these stars are no longer bound; the entire association expands. Many researchers, starting with \citet{L84}, have used N-body simulations to model this sudden release of the stars through flattening of the potential well.

\subsection{Mantle and Tide}

It is difficult to be more quantitative in the foregoing account of OB~associations, e.g., to estimate the number of clumps that supply most low-mass stars. The reason is that we do not really know how productive each clump is. The efficiency $\chi$ pertains only to the current epoch. But the picture offered here is that each clump extrudes stars even as it produces them. Thus, the true fraction of cloud mass turned into stars is greater than $\chi$. The mantle could, as in the illustrative examples, contain an equal number of stars as the cloud, and its total mass only grows in time.

This uncertainty also colors any discussion of T~associations and open clusters. The number of distinguishable clouds generating these groups can only be gleaned through observation of specific regions. To simplify matters, while capturing the essential physical ideas, I will focus on a single cloud. Figure~2 sketches a plausible chain of events for that object. This conjectured scenario, as yet wholly qualitative, addresses two longstanding questions. Why do older T~associations apparently disperse so quickly? Why is it that a group intermediate in mass between T and OB~associations remains gravitationally bound following cloud destruction?

\begin{figure}
    \hbox{\hspace{+0.00truecm}\includegraphics[width=\columnwidth]{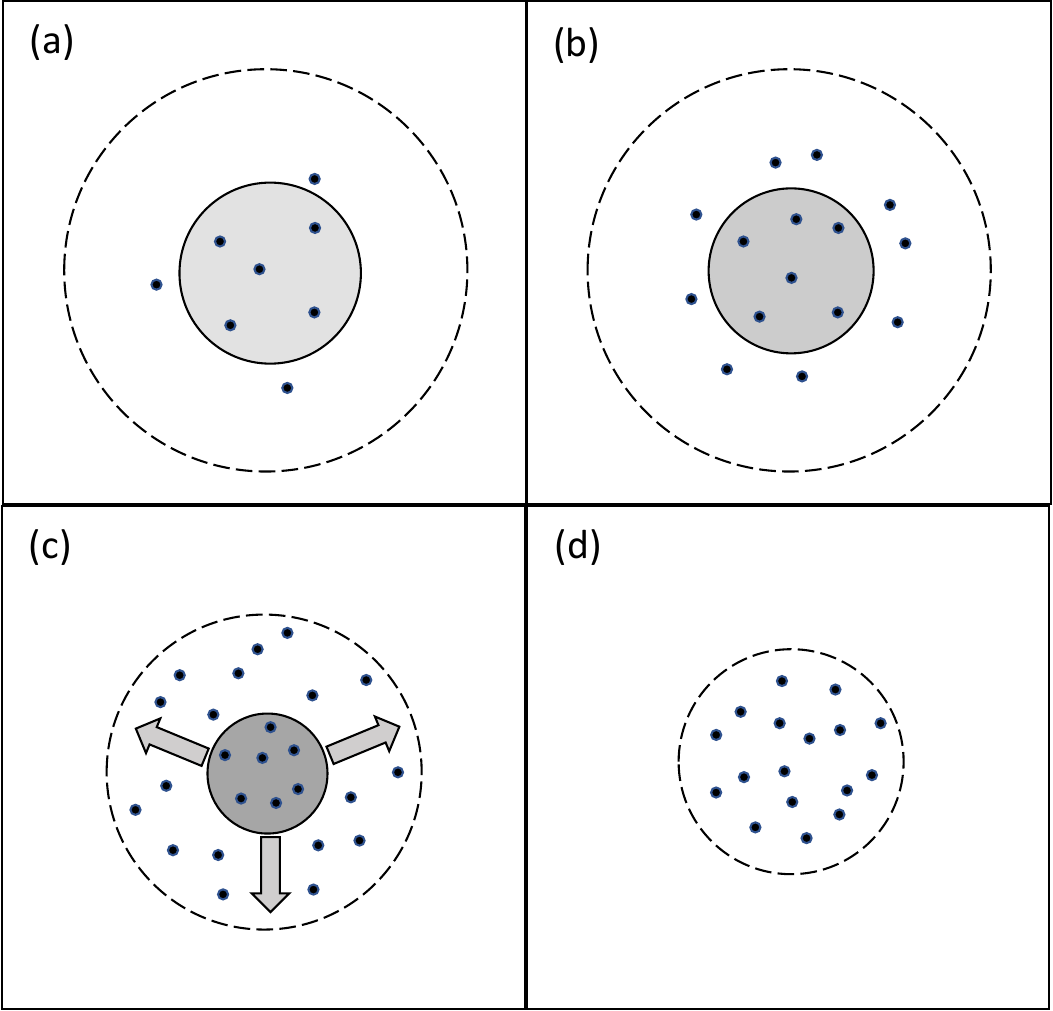}}
    \caption{Evolution of a star-forming gas cloud and its stellar mantle. (a) The cloud ({\it solid circle}), which is bounded by the external medium, forms stars; some drift into the mantle. Surrounding the mantle is the Galactic tidal radius ({\it dashed circle}). (b) The cloud shrinks and the mantle expands, but the tidal radius is unchanged. (c) Stars inside the cloud disperse gas, and the tidal radius shrinks to meet the mantle. (d) The cloud vanishes, leaving a stellar group that may or may not be gravitationally bound.} 
\end{figure}

This cloud, like all others, contracts in quasi-static fashion. As the density in the deep interior rises, so does the rate of star formation in that region. Some young stars drift out of the cloud and into the external medium, establishing the mantle (panel~a). There is no appreciable influence from the Galactic tidal field at this point, since $R_T$ is well removed from the system. As time passes. the cloud contracts further, creating more stars. These drift into the mantle, which itself expands. The tidal radius, however, remains fixed, since the mass interior to it has not changed (panel b).

There is net mass loss only when the cloud itself begins to disperse. How this dispersal occurs, in the absence of massive stars, is still not established. The most likely culprit is the energy and/or momentum injected by the stars being produced, in the form of stellar jets and molecular outflows. We cannot explore further the dispersal process, a subject of continued investigation by both observers and theorists. 
However, we note two points of interest, both related to the accelerating pace of star formation. Since jets and the outflows they produce are strongest soon after stellar birth, cloud dispersal should track the total star formation rate. The latter accelerates, so the time for dispersal is relatively brief, plausibly less than the initial crossing time. Secondly, the cloud's deep interior, which creates most stars, must continue to contract even as the total cloud mass diminishes. Gas ejection thus probably begins in the cloud's outer layers, and then works its way in.

Returning to the bigger picture, we note that $R_T$ depends weakly on the interior mass. Thus, a large fraction of the molecular gas must be shed before this radius shrinks appreciably. At some point, the outer edge of the mantle, $R_m$, meets $R_T$ (panel c). Now stars can also escape, as they cross the tidal boundary and enter the field population. The system eventually consists of a bare, stellar system whose outer boundary is the tidal radius (panel d).

The cloud gas that provided much of the self-gravity has effectively vanished, so the remnant stars are usually unbound. Outer members of the group cross $R_T$, lowering further the interior mass and causing $R_T$ to shrink more. This runaway process, ending in complete tidal disruption, is the ultimate fate of T~associations. The time spanned from the start of cloud dispersal to group annihilation must be relatively brief, as we see few examples of relatively populous associations consisting of older T~Tauri stars (see below for possible candidates). This observational fact is simply a restatement of the longstanding post-T Tauri problem, first enunciated by \citet{H78}.

It may happen that enough stars remain after the cloud disappears for the stellar system to be gravitationally bound. This would be the case if the parent cloud has a mass above the very lowest regime, but still not high enough to produce a massive star. The resulting system is a stable
open cluster. By virtue of its origin, the group is bound by the tidal radius, as is observed in real open clusters. This intermediate case must be relatively rare, since the inferred birthrate of open clusters in the Galaxy is the lowest of all three groups \citep[][]{M78}.

\section{Discussion and Conclusions}

The existence of stellar mantles, while not yet an established fact, is the natural outcome of a process that is observed, the drift of young stars from their birthsites. While the actual drift speed has not been measured directly, it may be inferred, at least approximately, from the progression of stellar evolutionary stages within each parent cloud. Stars which have drifted beyond the cloud boundary should be even more evolved. That is, we expect the mantle to consist of predominantly Class~II, and even Class~III, sources, the latter in its outer portion. Current observations of low-mass ``walkaway" stars (as opposed to the more familiar, massive runaways) in several star-forming regions are encouraging in this regard \citep{Z23}. Future studies with GAIA could identify convincingly this halo of stars and establish their drift velocity.

I have suggested that the physical origin of the drift lies in the overcrowding of stars as they form, and represents the outward diffusion of this excess population. This picture is quite different from the slingshot mechanism proposed by \citet{S18}. Here, the stars' parent filament contains a longitudinal magnetic field with helical topology. The field is oscillating by virtue of an MHD wave propagating along its length. As any segment of the filament reaches its maximum excursion and turns around, it flings off stars. While the basic idea of filament motion (regardless of its origin) is plausible, it remains true that stars form inside dense cores. The cores are integral parts of filaments, and are observed to gain mass inside them (recall \S~2.1). A star probably leaves its filament by partially disrupting it, dispersing the dense core via the same jets and outflows that ultimately decimate the entire parent cloud. In any case, future observations will establish whether existing filaments indeed harbor such longitudinal magnetic fields as the slingshot mechanism postulates.

It is interesting that even older pre-main~sequence stars are observed in the outskirts of many star-forming regions, some tens pf parsecs from the parent clouds. They were discovered three decades ago, by the X-ray satellite ROSAT. The origin of this ``distributed" population was the subject of much debate, and it is fair to say that no consensus prevailed \citep[for a review of the observations, see][]{N99}. These stars are too distant from the clouds to be their associated stellar mantles, which only expand subsonically. However, the evolutionary picture sketched in \S~4.2 suggests an alternative explanation. Most clouds forming T~associations are of the very lowest mass. Their internal stars destroy these clouds relatively quickly, whereupon the remnant stars are tidally dispersed over a wide area. Apart from the explicit role of the Galactic tidal field, this was the ``cloudlet" picture advocated by \citet{F96}, who also critically discussed a number of other, proposed models.

I note also the contrast between the account of open cluster formation proposed here and what has become the prevalent one. According to this view, open clusters are the bound remnants of expanding OB~associations. In the most widely cited illustrative simulation, \citet{K01} started with a cluster of 10,000 identical stars, embedded in a cloud of twice their total mass. The cloud was represented as an external potential well. After some interval, this potential was quickly flattened. About of a third of the stars remained as a bound cluster. There have subsequently been many variants of this calculation. 

The picture offered here is quite different. Massive stars play no role, and cloud dispersal, still assumed to be relatively brief, stems from low-mass stellar jets and outflows. Moreover, the parent cloud spawned, not a large surplus of stars, but about the same number observed in the present-day cluster, provided it is still relatively young. 

While this new picture remains to be quantified and tested, the currently popular view faces observational difficulties. The conjectured surplus stars are not seen, i.e., there are no embedded clusters with over a thousand members \citep{L03}. Even if they exist, a star formation efficiency of \hbox{$\chi = 0.33$} in the parent cloud is well above the observed range, as discussed above. Other simulations managed to obtain a bound cluster using a lower $\chi$, but required rather artificial assumptions: endowing the stars with a subvirial velocity dispersion \citep{V89}, postulating a high central stellar density \citep{G97}], or stretching out the duration of gas dispersal \citep{Ba07}. Lastly, the simulations typically find a radius for the bound system that is deep inside the corresponding tidal value \citep[e.g.,][Fig.~1]{K01}; this result is contrary to observations.

This last point bears repeating. A summary of my picture is that the Galactic tidal force erodes the expanding mantle to produce either a disintegrating T~association or a stable open cluster. Here, I have only motivated these ideas, including the existence of the mantle itself, in a largely qualitative manner. Exploring them through a direct simulation that includes both gas and stellar dynamics, even without the interstellar magnetic field, is not a realistic prospect. The best approach might be the one suggested here, 
a semi-analytic treatment in which stars and gas are two components of a single fluid. The dynamical equations derived in this paper can serve as the basis for this approach. Outward drift of the stars and production of the mantle should arise naturally, as both components adjust to maintain force balance in the quasi-statically contracting cloud.


\section*{Acknowledgements}

Early in this project, I benefited from extensive interaction with Eric Huff and Aaron Lee. More recently, Gaspard Duch{\^e}ne has offered critical feedback, while Philippe Andr\'e, Phil Myers and Nick Wright continue to educate me on observational matters. I thank the referee, whose comments resulted in an improved final version of this paper.

\section*{Data Availability}


No new data were generated or analyzed in support of this research.







\appendix

\section{Derivation of Dynamical Equations}


Let $f_g$ and $f_s$ be the distribution functions of gas and stars, respectively. That is, each is the 
number of particles (gas fluid elements and stars) per unit volume in six dimensional position-velocity 
phase space, where the elementary volume is $d^3 x\,d^3 v$. The amended Boltzmann equations are, in 
Cartesian coordinates,
\begin{align}
    {{\partial f_g}\over{\partial t}} \,+\, v_k\,{{\partial f_g}\over{\partial x_k}} \,-\, 
    {{\partial \psi}\over{\partial x_k}}\,{{\partial f_g}\over{\partial v_k}} \,&=\,
    \left({{\partial f_g}\over{\partial t}}\right)_c \,-\, {{s_\ast f_g}\over{\rho_g}} \\
    {{\partial f_s}\over{\partial t}} \,+\, v_k\,{{\partial f_s}\over{\partial x_k}} \,-\, 
    {{\partial \psi}\over{\partial x_k}}\,{{\partial f_s}\over{\partial v_k}} \,&=\,
    \left({{\partial f_s}\over{\partial t}}\right)_c \,+\, {{s_\ast f_s}\over{\rho_s}} \,\,.
\end{align}
Here, $\psi$ is the gravitational potential generated by both stars and gas, $\rho$ the mass density,
and $s_\ast$ the mass of stars being formed per volume of gas per time. The two righthand terms represent 
the effect of collisions, both within a species and between them, and the influence of star formation. 
These final terms were specifically chosen to yield the correct mass continuity equations.

Let $\chi_g$ and $\chi_s$ be quantities that are conserved in intraspecies collisions. After multiplying
equations~(A1) and (A2) by $\chi_g$ and $\chi_s$, respectively, and integrating over velocity, the collision
terms vanish, but those involving star formation do not. Momentarily surpressing the subscripts, we write
the result as
\begin{equation}
    {{\partial{\left(n\!<\!\!\chi\!\!>\right)}}\over{\partial t}} \,+\,
    {{\partial\left( n\!<\!\!\chi v_k\!\!> \right)}\over{\partial {x_k}}} \,+\,
    {{\partial\psi\over{\partial {x_k}}} n \!<\!\!{{\partial\chi}\over{\partial v_k}}}\!\!> 
    \,\pm\,{s_\ast\over m} \!<\!\!\chi\!\!> \,=\, 0\,\,,
\end{equation}
where the + and $-$ signs refer to gas and stars, respectively. Here, $m$ is the average mass of a gas fluid 
element or star, \hbox{$n\equiv \int\!\!f\,d^3v$} is the number density, and the brackets signify a weighted 
average over velocity, e.g., \hbox{$<\!\!\chi\!\!>\,\equiv \,n^{-1}\!\!\int\! \chi f \,d^3v$}. The effect of 
interpecies collisions is to enforce, through momentum and energy exchange, a common bulk speed and velocity 
dispersion.

We first set \hbox{$\chi_g = m_g$}. Inserting this quantity into equation~(A3), we also note that the full
velocity of each species is the sum of its bulk and fluctuating parts, \hbox{$v_k\,=\,u_k + w_k$}, but that only 
the bulk velocity survives after averaging: \hbox{$<\!\!v_k\!\!>\,=\,u_k$}. After identifying 
\hbox{$\rho_g = n_g\,m_g$}, we find the mass continuity equation for gas:
\begin{equation}
    {{\partial\rho_g}\over{\partial t}} \,+\, {{\partial(\rho_g u_k)}\over{\partial x_k}} \,+\,s_\ast\,=\,0\,\,,
\end{equation}
where the effect of star formation is manifest. Following the analogous procedure for stars yields 
\begin{equation}
    {{\partial\rho_s}\over{\partial t}} \,+\, {{\partial(\rho_s u_k)}\over{\partial x_k}} \,-\,s_\ast\,=\,0\,\,.
\end{equation}
The sum of equations~(A4) and (A5) is the total mass continuity equation:
\begin{equation}
    {{\partial\rho}\over{\partial t}} \,+\, {{\partial(\rho u_k)}\over{\partial x_k}} \,=\,0\,\,.
\end{equation}
In this last equation, \hbox{$\rho \equiv \rho_g + \rho_s$} signifies the total mass density. 

We next use, in equation~(A3), \hbox{$\left(\chi_g\right)_i = m_g\,v_i$}. While averaging $w_i$ gives 
zero, averaging the product, $w_i w_k$ does not. Rather, we obtain the gas pressure, via the relation
\hbox{$\,\,\rho_g\!<\!\!w_i\,w_k\!\!> = P_g \,\delta_{ik}$}. The gas momentum equation, in ``conservation
form," is
\begin{equation}
    {{\partial\!\left(\rho_g\,u_i\right)}\over{\partial t}} \,+\,
    {{\partial\left(\rho_g u_i u_k\right)}\over{\partial x_k}} \,+\,
    {{\partial P_g}\over{\partial x_i}}\,+\,
    \rho_g {{\partial\psi}\over{\partial x_i}} \,+\,s_\ast\,u_i \,=\, 0.
\end{equation}
Newly formed stars, since they are born with the same bulk velocity as the gas, naturally carry off 
momentum. However, there is no associated force on gas elements. To see this, we multiply the appropriate
mass continuity equation~(A4), by $u_i$, and subtract this equation from the previous one. We find
\begin{equation}
    \rho_g\,{{\partial u_i}\over{\partial t}} \,+\,
    \rho_g\,u_k\,{{\partial u_i}\over{\partial x_k}} \,+\,
    {{\partial P_g}\over{\partial x_i}}\,+\,
    \rho_g {{\partial\psi}\over{\partial x_i}} \,=\, 0 \,\,,
\end{equation}
which is just the Euler equation for the gas.

When we apply the same manipulations to the stars, we obtain the corresponding equation for this fluid:
\begin{equation}
    \rho_s\,{{\partial u_i}\over{\partial t}} \,+\,
    \rho_s\,u_k\,{{\partial u_i}\over{\partial x_k}} \,+\,
    {{\partial P_s}\over{\partial x_i}}\,+\,
    \rho_s {{\partial\psi}\over{\partial x_i}} \,=\, 0 \,\,.
\end{equation}
Here, \hbox{$P_s\,\equiv\,\rho_s\!<\!\!|w|^2\!\!>\!/3$} is the partial pressure supporting the 
stars against gravity. For both species, we may write \hbox{$\,P = \rho\,c^2$}, where $c$ is an 
effective sound speed, shared by gas and stars. We assume this quantity to be spatially constant, 
but to change with time. Another varying spatial constant shared by the two species is the specific
energy, \hbox{$\epsilon\,\equiv\,(3/2)\,c^2$}.

To explore global energy conservation, we start by multiplying the gas momentum equation~(A4), 
expressing mass conservation for gas, by $(1/2)\,u^2$. We next multiply the gas momentum
equation~(A8) by the velocity component $u_i$. Adding the two results gives the work equation for gas:
\begin{equation}
\begin{aligned}
    {{\partial{\phantom t}}\over{\partial t}}\!\left( {1\over 2} \rho_g u^2\ \right) \,+\,
    {{\partial{\phantom{x_k}}}\over{\partial x_k}}\!\left({1\over 2} \rho_g u^2 u_k\right) \,&+\,
    u_k{{\partial P_g}\over{\partial x_k}} \,+\, 
    \rho_g u_k {{\partial\psi}\over{\partial x_k}} \\
    \,&+\,{1\over 2} s_\ast \,u^2
    \,=\,0 \,\,.  
\end{aligned}
\end{equation}

We may derive a relation expressing total energy conservation in the gas by setting 
\hbox{$\chi_g\,=\,(1/2) m_g v^2$} in equation~(A3). This step requires justification. supplied in the text. The resulting total energy equation may be written
\begin{equation}
{{\partial{\phantom t}}\over{\partial t}}\!\left({1\over 2} \rho_g u^2 + \rho_g\epsilon\right) \,+\,
{{\partial F^g_k}\over{\partial x_k}} \,+\,\rho_g u_k {{\partial\psi}\over{\partial x_k}} \,+\,
{1\over 2} s_\ast u^2 \,+\, s_\ast \epsilon \,=\, 0 \,\,.
\end{equation}
The quantity $F^g_k$ is the total energy flux:
\begin{equation}
    F^g_k \,\equiv\, {1\over 2}\rho_g u^2 u_k \,+\,u_k P_g\,+\,\rho_g u_k \epsilon \,+\,
    F^g_{{\rm mech},k}  \,\,,
\end{equation}
where the last righthand term is the mechanical flux associated with the bulk transport of turbulence:
\begin{equation}
    F^g_{{\rm mech},k} \, = \,  {1\over 2}\,\rho_g \!<\! |w|^2 \,w_k\!> \,\,.
\end{equation}

A more useful relation is the internal energy equation for the gas. We derive this by subtracting 
the work equation~(A10) from the total energy equation~(A11). This gives
\begin{equation}
    {{\partial\!\left(\rho_g \epsilon\right)}\over{\partial t}} \,+\, 
    P_g {{\partial u_k}\over{\partial x_k}} \,+\,
    {{\partial\!\left(\rho_g \epsilon \,u_k\right)}\over{\partial x_k}} \,-\,
    s_\ast\,\epsilon \,+\,
    {{\partial F^g_{{\rm mech},k}}\over{\partial x_k}}
    \,=\,0 \,\,. \nonumber
\end{equation}
We account for dissipation by adding, to the mechanical flux, a radiative one. Thus, the full internal 
energy equation is 
\begin{equation}
    {{\partial\!\left(\rho_g \epsilon\right)}\over{\partial t}} + 
    P_g {{\partial u_k}\over{\partial x_k}} +
    {{\partial\!\left(\rho_g \epsilon \,u_k\right)}\over{\partial x_k}} -
    s_\ast\,\epsilon 
    +\,{{\partial\!\left( F^g_{{\rm mech},k} + F^g_{{\rm rad},k}\right)}\over{\partial x_k}}
    \,=\,0 \,\,. 
\end{equation}

Turning to the stars, the work equation is identical to that for the gas, except for the usual
sign change in the star formation term:
\begin{equation}
\begin{aligned}
    {{\partial{\phantom t}}\over{\partial t}}\!\left( {1\over 2} \rho_s u^2\ \right) \,+\,
    {{\partial{\phantom{x_k}}}\over{\partial x_k}}\!\left({1\over 2} \rho_s u^2 u_k\right) \,&+\,
    u_k{{\partial P_s}\over{\partial x_k}} \,+\, 
    \rho_s u_k {{\partial\psi}\over{\partial x_k}} \\
    \,&-\,{1\over 2} s_\ast\,u^2  \,=\,0 \,\,.  
\end{aligned}
\end{equation}
The total energy equation is also similar:
\begin{equation}
{{\partial{\phantom t}}\over{\partial t}}\!\left({1\over 2} \rho_s u^2 + \rho_s\epsilon\right) \,+\,
{{\partial F^s_k}\over{\partial x_k}} \,+\,\rho_s u_k {{\partial\psi}\over{\partial x_k}} \,-\,
{1\over 2} s_\ast u^2 \,-\, s_\ast \epsilon \,=\, 0 \,\,,
\end{equation}
The energy flux in the stars is now
\begin{equation}
    F^s_k \,\equiv\, {1\over 2}\rho_s u^2 u_k \,+\,u_k P_s\,+\,\rho_s u_s \epsilon \,+\,
    F^s_{{\rm mech},k}  \,\,,
\end{equation}
where
\begin{equation}
    F^s_{{\rm mech},k} \, = \,  {1\over 2}\,\rho_s \!<\! |w|^2 \,w_k\!> \,\,.
\end{equation}
Since the stellar motion is not dissipative, the internal energy equation for the stars is 
\begin{equation}
    {{\partial\!\left(\rho_s \epsilon\right)}\over{\partial t}} + 
    P_s {{\partial u_k}\over{\partial x_k}} +
    {{\partial\!\left(\rho_s \epsilon \,u_k\right)}\over{\partial x_k}} 
    + s_\ast\,\epsilon 
    +\,{{\partial\!\left( F^s_{{\rm mech},k}\right)}\over{\partial x_k}}
    \,=\,0 \,\,. 
\end{equation}

The total internal energy equation. which we obtain by summing equations~(A4) and (A19),
no longer contains the star formation rate:
\begin{equation}
    {{\partial\!\left(\rho \epsilon\right)}\over{\partial t}} \,+\, 
    P \,{{\partial u_k}\over{\partial x_k}} \,+\,
    {{\partial\!\left(\rho \,\epsilon \,u_k\right)}\over{\partial x_k}} \,+\,
    {{\partial F_k}\over {\partial x_k}} = 0 \,\,.
\end{equation}
Here, the total energy flux in both gas and stars is \hbox{$\,F_k\equiv\,F^g_{{\rm mech},k} + 
F^s_{{\rm mech},k} + F^g_{\rm rad}$}. We may rewrite this last equation in a more compact,
physically appealing form. Returning to the full mass continuity equation~(A6), we multiply 
it by the specific enthalpy, \hbox{$h \equiv \epsilon + P/\rho$}, and subtract the result
from equation~(A20). After restoring the expression for $\epsilon$, we find
\begin{equation}
    \rho\,c^2 \,{{D\phantom{t}}\over{Dt}}\,{\rm ln}\left({c^3\over\rho}\right) \,+\,
    {{\partial F_k}\over{\partial x_k}} \,=\,0 \,\,.
\end{equation}
We have employed here the convective time derivative:
\begin{equation}
    {D\phantom{t}\over{Dt}} \,\equiv\,{\partial\phantom{t}\over{\partial t}} \,+\,
    u_k {\partial\phantom{x_k}\over{\partial x_k}} \,\,.
\end{equation}
The logarithmic quantity is the specific entropy of the cloud, and we may term (A21)
the ``heat equation," after the analogous relationship in stellar structure theory.



\bsp	
\label{lastpage}
\end{document}